\documentclass{PoS}

\usepackage{graphics}
\usepackage{amssymb}
\usepackage{amsmath}

\title{Indirect dark matter search with the ANTARES neutrino telescope}

\ShortTitle{Indirect Dark Matter search in the Sun with ANTARES}

\author{\speaker{Guillaume Lambard}\thanks{on behalf of the ANTARES collaboration}\\
        IFIC - Instituto de F\`{i}sica Corpuscular - Edificio Institutos de Investigaci\'{o}n, 
        Apartado de Correos 22085 E-46071 Valencia, Spain\\
        E-mail: \email{lambard@ific.uv.es}}

\abstract{Using the data recorded by the ANTARES neutrino telescope during      
2007 and 2008, a search for high energy neutrinos coming from the     
direction of the Sun has been performed. The neutrino selection       
criteria have been chosen so as to maximize the rejection of the      
atmospheric background with respect to possible signals produced by   
the self-annihilation of weakly interactive massive particles	      
accumulated in the centre of the Sun. After data unblinding, the      
number of neutrinos observed was found to be compatible with	      
background expectations.  The results obtained were compared to the   
fluxes predicted by the Constrained Minimal Supersymmetric Standard   
Model, and 90\% upper limits for this model were obtained. Our	      
limits are competitive with those obtained by other neutrino	      
telescopes such as IceCube and SuperKamiokande, which give ANTARES limits for  
the spin-dependent WIMP-proton cross-section that are more stringent  
than those obtained by direct search experiments.}

\FullConference{VIII International Workshop on the Dark Side of the Universe,\\
		June 10-15, 2012\\
		Rio de Janeiro, Brazil}

\begin{document}

\section{Introduction}
\label{introduction}

There is compelling evidence that $83$\% of the matter in the Universe 
is in the form of non-baryonic and non-relativistic matter that does interact 
weakly with the ordinary matter, the so-called dark matter. Much of this 
evidence comes from its gravitational effects on the motion of galaxies, 
clusters of galaxies and from the large scale structure of the Universe. 
The existence of dark matter is a key component of our present standard 
cosmological model, and results from the study of the CMB anisotropies 
and gravitational lensing in galaxy clusters further support its existence 
(for a review of the evidence, candidates and constraints, 
see~\cite{darkmatter} and~\cite{pdgdm}).

One of the most favoured hypothesis is that dark matter is made of
weakly interacting massive particles (WIMPs) that are embedded in the
visible, baryonic part of the galaxies and surround them in the form
of a halo. There are a variety of candidates for WIMPs, among which
those provided by theories based on supersymmetry (SUSY) or universal
extra dimensions (UED) attract a great deal of interest. In some
classes of SUSY, the lightest particle is stable due to the conservation 
of R-parity that forbids its decay to standard particles, making it a 
good candidate as a dark matter WIMP.

In addition to its gravitational effects, the search for evidence of
the existence of WIMPs is at present performed, on the one hand, by
looking for the recoiling products of the elastic scattering of dark
matter particles off normal baryonic matter in suitable detectors, the
so-called direct searches, and on the other hand, indirectly by the
observation of the final products of the possible annihilation of
WIMPs that have accumulated in astrophysical objects. WIMPs can
scatter elastically in the Sun or the Earth and become trapped in
their gravitational potential wells, accumulating in sizeable numbers
over the age of the solar system, therefore a very wide region in the Galaxy must
have contributed, reducing the dependence on the detailed structure of the dark 
matter halo. The WIMPs proposed by SUSY are Majorana fermions that 
can self-annihilate, giving rise to standard particles. Neutrinos will be at 
the end of the decay chain of the products of the WIMP self-annihilation, 
they can escape these astrophysical objects and be detected by neutrino telescopes 
on the Earth.

In this paper, the indirect search for dark matter looking
for high energy neutrinos coming from the Sun using the data recorded
by the ANTARES neutrino telescope in 2007 and 2008 is reported. The number of
neutrinos observed is compared to the neutrino fluxes predicted by
the Constrained Minimal Supersymmetric Standard Model (CMSSM)~\cite{ellisintro}, a minimal 
supersymmetric extension of the Standard Model with supersymmetry-breaking 
scalar and gaugino masses constrained to be universal at the GUT scale.

The layout of the paper is as follows. In Section~\ref{antares}, the main features 
of the ANTARES neutrino telescope and the reconstruction algorithm used in this work 
are reviewed. In Section \ref{simulation}, the Monte Carlo simulations of 
the signal, expected from the investigated WIMP models, and the background, expected
from atmospheric muons and neutrinos, are described. In Section~\ref{optimisation},
the method used to optimise the criteria to select the
sample of neutrino events and reduce the background following a blind
data strategy is presented. Finally, the results obtained are discussed in
Section~\ref{results}, where limits on the models investigated are
imposed from the absence of a dark matter signal.


\section{The ANTARES neutrino telescope}
\label{antares}

ANTARES is the first undersea neutrino telescope and the largest of
its kind in the Northern Hemisphere~\cite{antares}. It is located at
$2475$~m depth in the Mediterranean Sea, $40$ km offshore from Toulon
(France) at $42^{\circ} 48$' N and $6^{\circ} 10$' E.

The telescope consists of $12$ mooring detection lines made up of 25
storeys each. The standard storey is composed of a local control
module that contains the front-end and slow-control electronics
and three optical modules (OMs) that house a 10-inch photomultiplier. 
The OMs are looking 45$^{\circ}$ downwards in order to optimise their acceptance to
upgoing light and to avoid the effect of sedimentation and biofouling. The length 
of the lines is 450 m and the horizontal distance between neighbouring lines is 60-75 m.  
The absolute time accuracy is ensured at the millisecond level by the UTC
time provided by the GPS system connected to the clock system of the
detector. The relative time between the elements of the
detector is achieved by calibration of the lines in the laboratory 
with short light pulses, by the use of a 25 MHz clock system and by
the operation of a series of optical beacons distributed along the
lines that emit short light pulses through the water. 
A relative timing of the order of one nanosecond is reached. 
Additional information on the detector can be found in reference~\cite{antares}.

Neutrinos are detected through the products of their interaction with
the matter inside or close to the detector. In the channel of neutrino
observation used in this work, a high energy neutrino interacts in the
rock below the detector producing a relativistic muon that can travel
hundreds of meters and cross the detector or pass nearby. This muon
induces Cherenkov light when travelling through the water, which is
detected by the OMs. From the time and position information of the
hits in the OMs, the direction of the muon -- which at high energy is
essentially that of the neutrino -- can be reconstructed.

Data-taking started in 2007 when only 5 lines of the detector where installed. 
The full detector was connected in May 2008 and has been operating ever
since, except for some periods in which repair and maintenance
operations have taken place. For this work the data taken during 2007
and 2008 has been used. Results of other searches using this
data-taking period can be found elsewhere~\cite{diffuse,ps,monopole}.

\bigskip
The reconstruction of the track from the position and time of the hits
of the Cherenkov photons in the OMs is a key ingredient of the physics
analysis. In this work, a dedicated fast algorithm to reconstruct the
muon track is used~\cite{bbfit}. The algorithm is based on the
minimization of a $\chi^{2}$-like quality parameter, $Q$, that uses
the differences of the expected and actual times of the detected
photons corrected by the effect of light absorption in water.

Monte Carlo simulations indicate that selecting tracks with a quality
parameter per degree of freedom of the fit smaller than $1.4$ results
in a purity of $90$\% for upward reconstructed multi-line atmospheric
neutrinos, the contamination being misreconstructed downgoing
atmospheric muon events, with an angular resolution of about few degrees 
at energies of tens of GeV, driven at $\sim40$\% by the kinematic 
of the neutrinos in low energy regime.

\section{Signal and background simulation}
\label{simulation}

The number of muon neutrinos as a function of their energy arriving at the
Earth's surface from the Sun's core, $dN_{\nu}/dE_{\nu}$, is computed
using the software package WimpSim~\cite{wimpsim}. The usual
self-annihilation channels ($q\bar{q}$, $l\bar{l}$, $WW$, $ZZ$, Higgs doublets
$\phi\phi^{*}$ and $\nu\bar{\nu}$) were simulated for $17$ different
WIMP masses from $10$~GeV to $10$~TeV. Oscillations between the three neutrino 
flavours both in the Sun and during their flight to Earth as well as 
$\nu$ absorption and $\tau$ lepton regeneration in the Sun are taken into 
account.

For the CMSSM, three main self-annihilation channels were chosen for
the lightest neutralino, $\tilde{\chi}_{1}^{0}$, namely: a soft
neutrino channel, $\tilde{\chi}_{1}^{0}\tilde{\chi}_{1}^{0}
\rightarrow b\bar{b}$, and two hard neutrino channels,
$\tilde{\chi}_{1}^{0}\tilde{\chi}_{1}^{0} \rightarrow W^{+}W^{-}$ and
$\tilde{\chi}_{1}^{0}\tilde{\chi}_{1}^{0} \rightarrow
\tau^{+}\tau^{-}$. Since which of these three channels is dominant
depends on the region of the CMSSM parameter space being
analysed~\cite{desai}, a 100\% branching ratio was assumed for 
all of them in order to explore them on an equal footing.

%

\bigskip
The main backgrounds for this search are muons and neutrinos produced in the 
interaction of cosmic rays with the atmosphere. Downgoing atmospheric muons 
dominate the trigger rate, which ranges from $3$ to$10$ Hz depending on the 
exact trigger conditions. They are simulated using 
Corsika~\cite{corsika}. Upgoing atmospheric neutrinos, which are recorded at a 
rate of $\sim$1 mHz (about four per day)~\cite{antares}, are simulated according 
to the parametrisation of the atmospheric $\nu_{\mu}$ flux from~\cite{bartol} 
in the energy range from $10$ GeV to $10$ PeV.

The Cherenkov light produced in the vicinity of the detector is 
propagated taking into account light absorption and scattering in sea
water. The angular acceptance, quantum efficiency
and other characteristics of the PMTs are taken from~\cite{amram} and
the overall geometry corresponded to the layout of the ANTARES
detector~\cite{antares} according to the data taking periods (from $5$ 
to $12$ line configurations).



\section{Optimisation of the event selection criteria}
\label{optimisation}

The data set used in this analysis comprises a total of $2,693$ runs
recorded between the $27^{th}$ of January 2007 and the $31^{st}$ of
December 2008, corresponding to a total livetime of $\sim$294.6
days. The detector consisted of $5$ lines for most of $2007$ and of
$9$, $10$ and $12$ lines for $2008$.

Only upgoing events are kept in the analysis. The muon tracks are 
required to have cos$\, \theta<$ 0.9998 in order to exclude those for 
which the fit stopped at the boundary. The fit is required to use a
number of hits greater than five in at least two lines in order to
ensure a non-degenerate 5-parameter fit with a proper reconstruction
of the azimuth angle.

The UTC time of the events is uniformly randomised on the period of the 
data taking in order to estimate the background in the Sun's direction from 
the data itself. The local coordinates ($\theta$, $\phi$) are kept so as 
to preserve the detector geometry in the optimisation of the selection criteria. 
This procedure provides a means to follow a {\it data blinding} strategy while 
using all the relevant information on the detector's performance.

The values used in the event selection criteria for the quality
parameter, $Q$, and for the aperture of the search cone in the Sun's
direction, $\Psi$, are optimised following the model rejection factor (MRF) 
technique~\cite{mrf}. For each WIMP mass and each annihilation channel, 
the values of $Q$ and $\Psi$ used are those that optimised the average upper 
limit on the $\nu_{\mu}+\bar{\nu}_{\mu}$ flux, $\bar{\phi}_{\nu_{\mu}+\bar{\nu}_\mu}$, 
as defined by:

\begin{equation}
\bar{\phi}_{\nu_{\mu}+\bar{\nu}_\mu} = \frac{\bar{\mu}^{90\%}}{\sum_{i} A_{eff}^{i}(M_{\rm WIMP}) \times T_{eff}^{i}}  \, ,
\label{mrfeq}
\end{equation}

\noindent where the index $i$ denotes the different periods in the detector infrastructure 
(5, 9, 10 and 12 detection lines), $\bar{\mu}^{90\%}$ is the average upper limit at
$90$\% confidence level (CL) computed from the time-scrambled data set and
using a Poisson distribution in the Feldman-Cousins approach~\cite{feldmancousins}; 
$T_{eff}^{i}$ is the livetime for each detector configuration in $2007$-$2008$, 
namely: $\sim$134.6, $\sim$38, $\sim$39 and $\sim$83 days for $5$, $9$, $10$ and $12$ 
lines respectively. The effective area averaged over the neutrino energy, 
$A_{eff}(M_{\rm WIMP})^{i}$, is defined as (the index $i$ is implied):

\begin{equation}
A_{eff}(M_{\rm WIMP}) =
\sum_{\nu,\bar{\nu}} \left ( \frac{\int_{E_{\nu}^{th}}^{M_{\rm WIMP}} A_{eff}(E_{\nu,\bar{\nu}}) \, \frac{dN_{\nu,\bar{\nu}}}{dE_{\nu,\bar{\nu}}} dE_{\nu,\bar{\nu}}}
{\int_{0}^{M_{\rm WIMP}}\frac{dN_{\nu}}{dE_{\nu}} dE_{\nu} \,+\, \frac{dN_{\bar{\nu}}}{dE_{\bar{\nu}}} dE_{\bar{\nu}}} \right )  \, ,
\label{aeffeq}
\end{equation}

\noindent where $E_{\nu}^{th}=10$~GeV is the energy threshold for
neutrino detection in ANTARES, $M_{\rm WIMP}$ is the WIMP mass, 
$dN_{\nu,\bar{\nu}}/dE_{\nu,\bar{\nu}}$ is the energy spectrum of the
neutrinos or the anti-neutrinos at the surface of the Earth for the three 
channels of interest in this analysis, and $A_{eff}(E_{\nu,\bar{\nu}})$ is 
the effective area of ANTARES as a function of the neutrino or anti-neutrino 
energy. Due to their different cross-sections these two effective areas are 
slightly different and therefore are studied separately.

%

\begin{figure}[!b]
\begin{minipage}[c]{.52\linewidth}
\includegraphics[width=\linewidth]{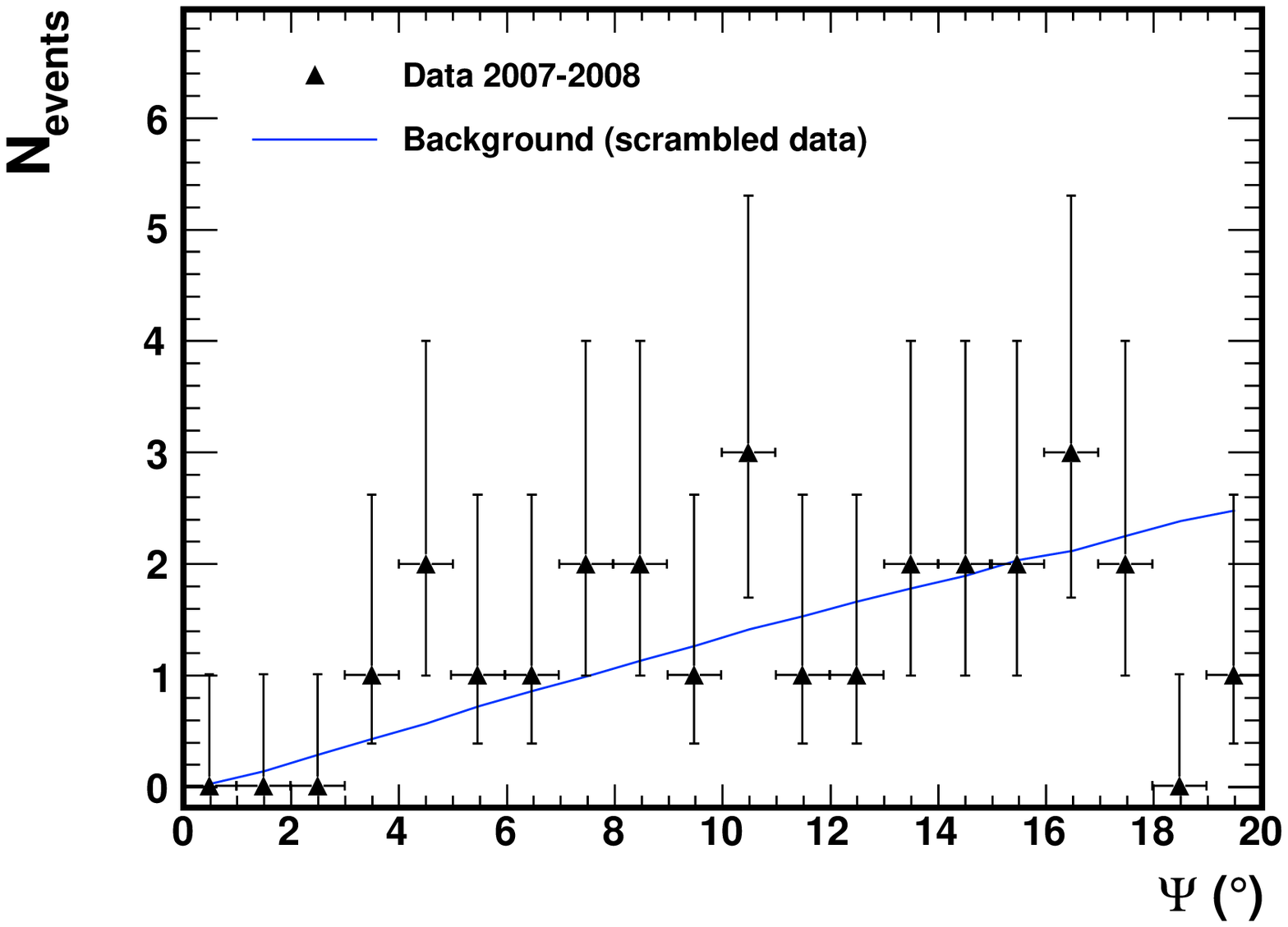}
\end{minipage}
\begin{minipage}[c]{.52\linewidth}
\includegraphics[width=\linewidth]{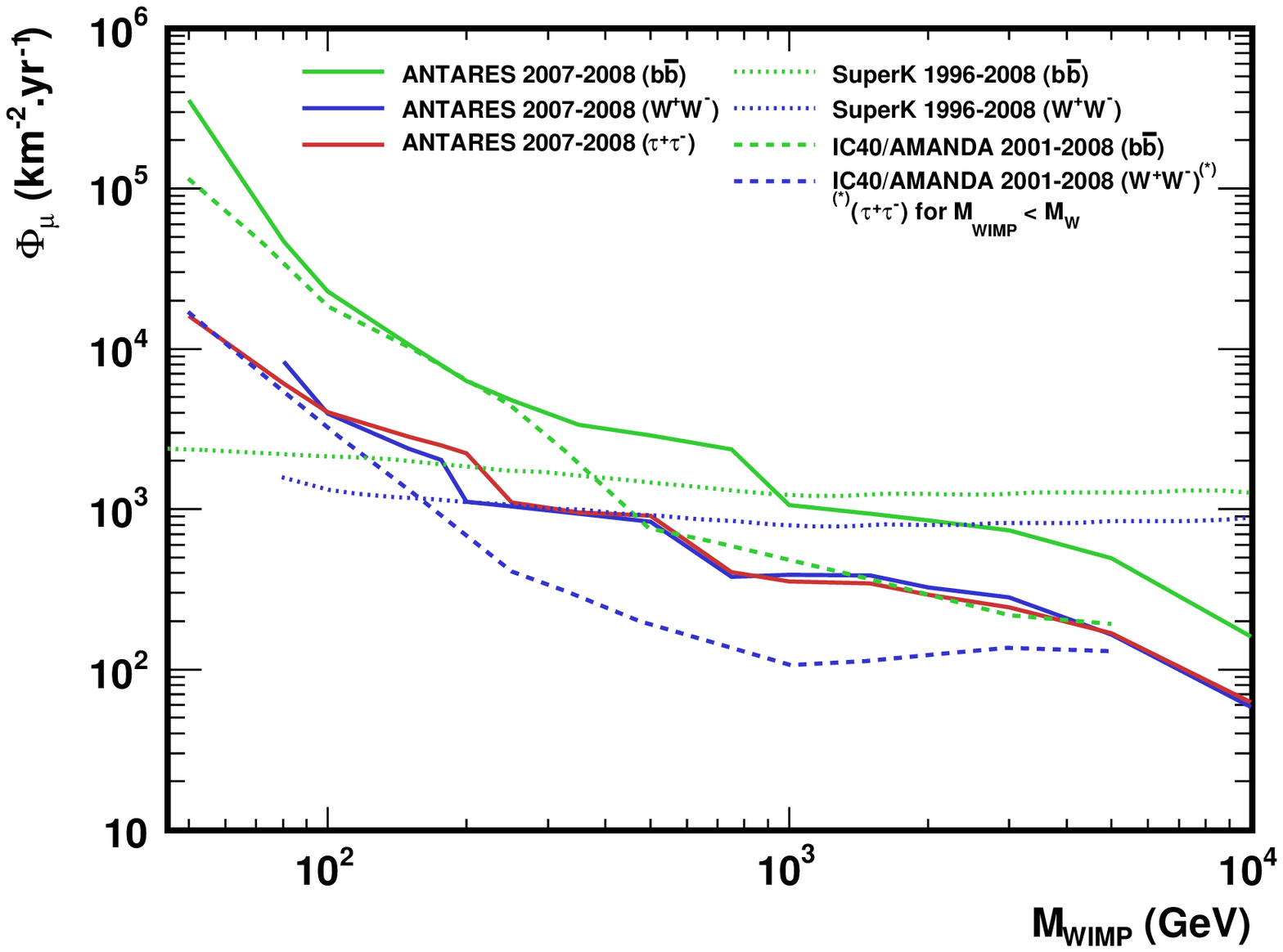}
\end{minipage}
\caption{{\bf Left:} Distribution of the spatial angle $\Psi
  \in$[$0^{o}$,$20^{o}$] of the event tracks with respect to the Sun's
  diretion for the expected background computed from the time-scrambled
  data (solid blue line) compared to the data after the basic
  selection criteria (black triangles) . A $1\sigma$ Poisson
  uncertainty is shown for each data point (black cross). {\bf Right:} $90$\% CL upper limit 
on the muon flux as a function of the WIMP mass in the range $M_{\rm WIMP}\in$[$50$ GeV;$10$ TeV] 
for the three channels $b\bar{b}$ (green), $W^{+}W^{-}$ (blue) and $\tau^{+}\tau^{-}$ (red, ANTARES only). 
The results from SuperKamiokande $1996-2008$~\cite{superk} (dotted lines) and IceCube-$40$ plus AMANDA 
$2001$-$2008$~\cite{icecube} (dashed lines) are also shown.}
\label{databkgfig}
\end{figure}


\section{Results and discussion}
\label{results}

Once the optimised values of $Q$ and $\Psi$ were obtained using the
time-scrambled data, the data sample is unblinded. Figure~\ref{databkgfig} 
shows the distribution of the spatial angle between the tracks of the events 
and the Sun's direction obtained after applying the basic selection criteria 
on the zenith angle and the minimum number of hits and lines. A total of $27$ 
events were found within a $20$ degrees spatial angle and no excess in the Sun's 
direction above the scrambled background are observed.

Using the values for the cuts obtained in the optimisation procedure,
limits on the $\nu_{\mu}+\bar{\nu}_{\mu}$ flux, $\phi_{\nu_{\mu}+\bar{\nu}_\mu}$, 
can be extracted from the data according to:

\begin{equation}
\phi_{\nu_{\mu}+\bar{\nu}_\mu} = \frac{\mu^{90\%}}{\sum_{i} A_{eff}^{i}(M_{\rm WIMP}) \times T_{eff}^{i}}  \, ,
\label{nulimiteq}
\end{equation}

\noindent where $\mu^{90\%}$ is the upper limit at $90$\% CL on the number of 
observed events and the rest of variables have the same meaning as in Eq.~\ref{mrfeq}.  

The corresponding limits for muons are calculated using a conversion factor between the neutrino 
and the muon fluxes ($\phi_{\mu}=\Gamma_{\nu\rightarrow\mu} \times \phi_{\nu_{\mu}+\bar{\nu}_\mu}$) 
computed using DarkSusy~\cite{gondolo}. Figure~\ref{databkgfig} (on the right) shows the 
90\% CL muon flux limits $\phi_{\mu}$ for the channels $b\bar{b}$, $W^{+}W^{-}$ and 
$\tau^{+}\tau^{-}$. The latest results from SuperKamiokande~\cite{superk} and 
IceCube-$40$ plus AMANDA~\cite{icecube} are also shown for comparison. Despite its smaller 
detector volume, the ANTARES limits both in the soft and hard channels are similar to those 
obtained by IceCube40 and AMANDA in the mass range $M_{\rm WIMP}\in$[$50$;$100$] GeV. In this mass 
range most of the sensitivity of the South Pole telescopes comes from AMANDA-II, 
whose effective area and energy threshold were similar to those of ANTARES in its 12-line 
configuration. Compared to the limits set by SuperKamiokande, those of
ANTARES are more stringent in the high mass region $M_{\rm WIMP} \,>\, 150$ GeV.

Systematic uncertainties have been taken into account and included in the 
evaluation of the limits using the approach of reference~\cite{conrad} by 
means of the {\tt Pole} software. The total systematic uncertainty on the detector efficiency
is around 20\% and comes mainly from the efficiency and time
resolution uncertainties of the OMs, the total angular resolution and
the absolute pointing accuracy.  This uncertainty translates into an
increase of the upper limit between 3\% and 6\% depending on the WIMP
mass.

%


\begin{figure*}[!h]
\begin{minipage}[c]{.52\linewidth}
\includegraphics[width=\linewidth]{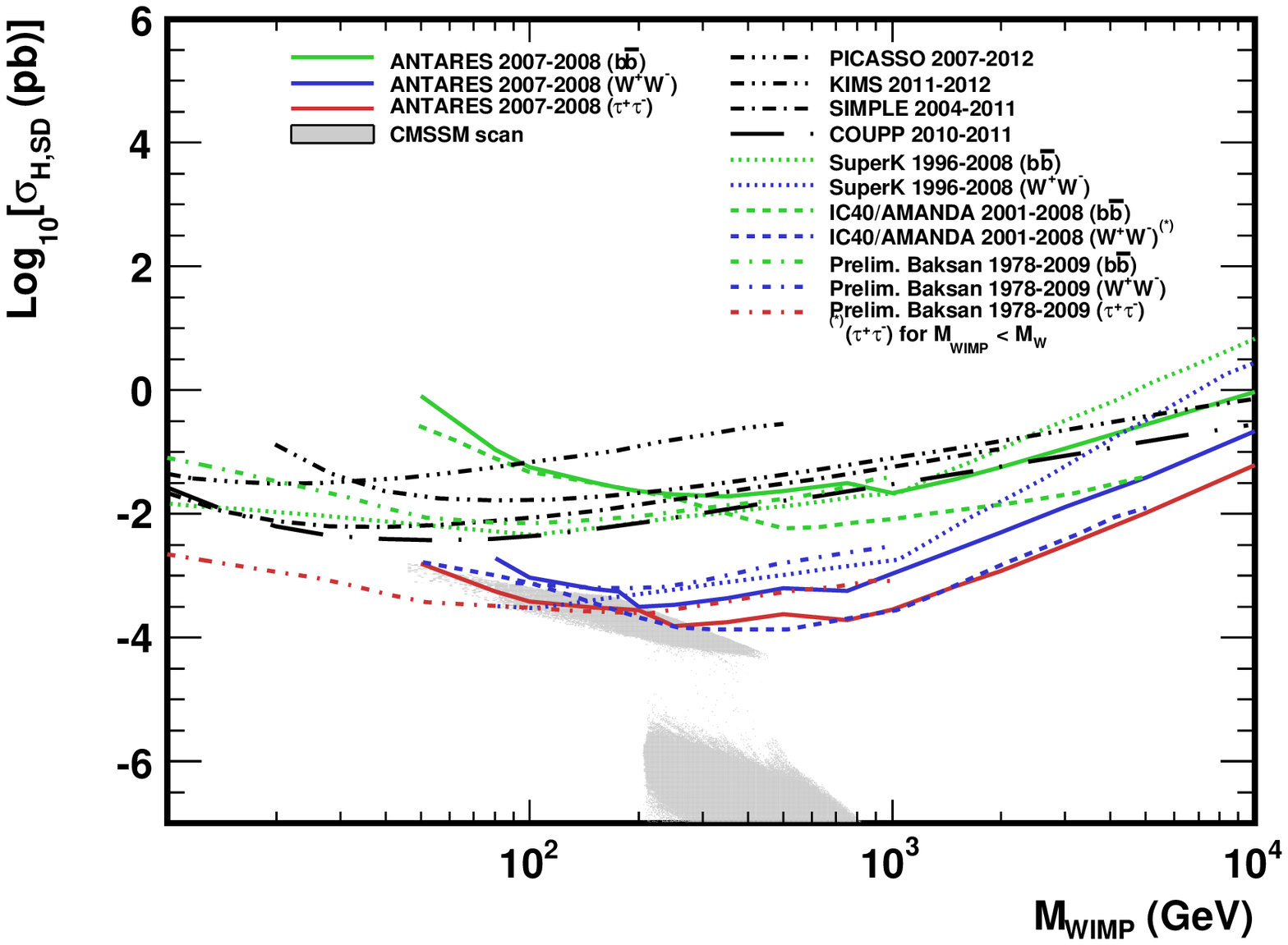}
\end{minipage}
\begin{minipage}[c]{.52\linewidth}
\includegraphics[width=\linewidth]{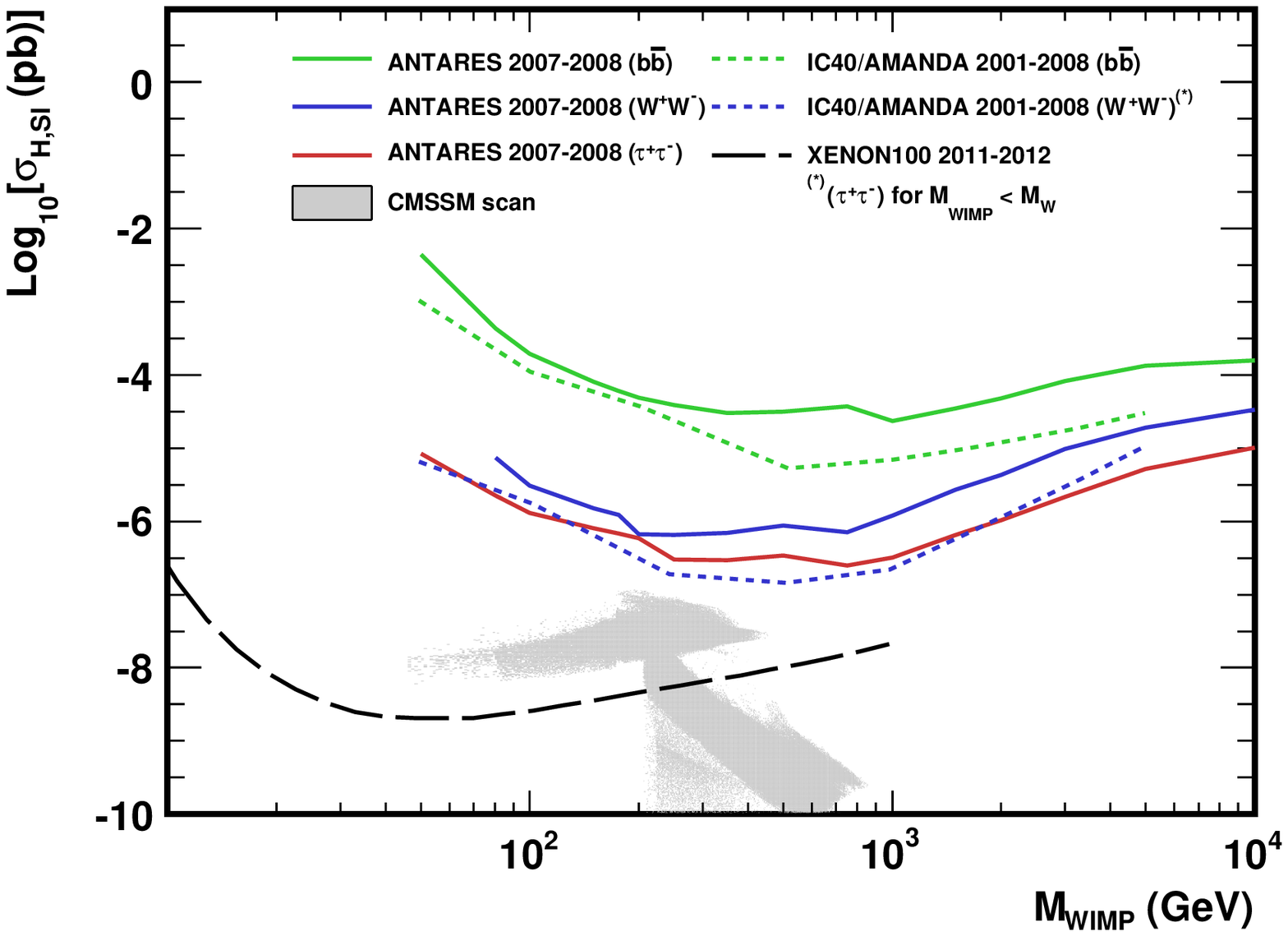}
\end{minipage}
\caption{$90$\% CL upper limits on the SD and SI WIMP-proton cross-sections 
(left and right hand respectively) as a function of the WIMP mass in the range 
$M_{\rm WIMP}\in$[$50$ GeV;$10$ TeV], for the three channels: $b\bar{b}$ (green), 
$W^{+}W^{-}$ (blue) and $\tau^{+}\tau^{-}$ (red), for ANTARES (solid line) compared to the 
results of other indirect search experiments: SuperKamiokande $1996-2008$~\cite{superk} (dotted lines) 
,IceCube-$40$ plus AMANDA $2001$-$2008$~\cite{icecube} (dashed lines) and Baksan $1978-2009$~\cite{baksan} 
(colored dot-dashed line) and the result of the most stringent direct search experiments (black): 
PICASSO $2007-2012$~\cite{picasso} (three-dots-dashed line), KIMS $2011-2012$~\cite{kims} 
(two-dots-dashed line), SIMPLE $2004-2011$~\cite{simple} (one-dot-dashed line), COUPP $2010-2011$~\cite{coupp} 
(one-dot-large-dashed line) and XENON100 $2011-2012$~\cite{xenon} (dashed line). For a comparison to the 
theoretical expectations, grid scans corresponding to the CMSSM model have been added.}
\label{sdsilimitcmssmfig}
\end{figure*}

\bigskip
Assuming equilibrium between the WIMP capture and
self-annihilation rates in the Sun, the limits on the spin-dependent (SD) and the
spin-independent (SI) WIMP-proton scattering cross-sections can be
obtained for the case in which one or the other is dominant.

A conservative approach to the dark matter local halo was considered
assuming a gravitational effect of Jupiter on the Sun's capture rate,
which reduces about $13$\%($87$\%) the SD capture and about $1$\%
($12$\%) the SI capture for $1$($10$) TeV WIMPs. A local dark matter
density of about $0.3$ GeV.cm$^{-3}$ was assumed. No additional dark
matter disk that could enhance the local dark matter density was
considered (see~\cite{edsjosdcs} for a discussion).

\bigskip
The $90$\% CL limits for the SD and SI WIMP-proton cross-sections extracted 
from the channels $b\bar{b}$, $W^{+}W^{-}$ and $\tau^{+}\tau^{-}$ are presented 
in Figure~\ref{sdsilimitcmssmfig}. The latest results from SuperKamiokande, 
IceCube-$40$+AMANDA and Baksan together with the latest and the most stringent limits from the 
direct search experiments PICASSO, KIMS, SIMPLE, COUPP and XENON100 are also shown. The allowed 
parameter space from CMSSM model according to the results from an adaptative grid scan 
performed with DarkSUSY~\cite{gondolo} are also shown. For a proper comparison 
all the limits presented in Figure~\ref{sdsilimitcmssmfig} are computed with a muon 
energy threshold at $E_\mu = 1$ GeV. For these figures, the shaded regions show a grid scan 
of the model parameter space taking into account the last limits for the Higgs boson mass 
from ATLAS and CMS merged together such as $M_{H} = 125 \pm 2 GeV$~\cite{buchmeller}. 
A relatively large constraint on the neutralino relic density $0 < \Omega_{CDM}h^{2} < 0.1232$ 
is used not to enclose the studied dark matter particle to only one possible nature.

\bigskip
The neutrino flux due to WIMP annihilation in the Sun is highly      
dependent on the capture rate of WIMPs in the core of the Sun, which 
in turn is dominated by the SD WIMP-proton cross-section. This makes 
these indirect searches better compared to direct search   
experiments such as KIMS and COUPP.  This is			     
not the case for the SI WIMP-proton cross-section, where the limits  
coming from direct search experiments such as CDMS and  	     
XENON100 are better thanks to their target			     
materials. Therefore, there is a sort of complementarity between both
types of searches.						     

\section*{Acknowledgments}
The authors acknowledge the financial support of the funding agencies:
Centre National de la Recherche Scientifique (CNRS), Commissariat \`a
l'\'ene\-gie atomique et aux \'energies alternatives (CEA), Agence
National de la Recherche (ANR), Commission Europ\'eenne (FEDER fund
and Marie Curie Program), R\'egion Alsace (contrat CPER), R\'egion
Provence-Alpes-C\^ote d'Azur, D\'e\-par\-tement du Var and Ville de La
Seyne-sur-Mer, France; Bundesministerium f\"ur Bildung und Forschung
(BMBF), Germany; Istituto Nazionale di Fisica Nucleare (INFN), Italy;
Stichting voor Fundamenteel Onderzoek der Materie (FOM), Nederlandse
organisatie voor Wetenschappelijk Onderzoek (NWO), the Netherlands;
Council of the President of the Russian Federation for young
scientists and leading scientific schools supporting grants, Russia;
National Authority for Scientific Research (ANCS), Romania; Ministerio
de Ciencia e Innovaci\'on (MICINN), Prometeo of Generalitat Valenciana
and MultiDark, Spain; Agence de l'Oriental and CNRST, Morocco. We also
acknowledge the technical support of Ifremer, AIM and Foselev Marine
for the sea operation and the CC-IN2P3 for the computing facilities.


\begin{thebibliography}{99}


\bibitem{darkmatter}G. Bertone, D. Hooper, J. Silk, Phys.Rept., 2005, {\bf 405}: pp. 279-390.

\bibitem{pdgdm}Particle Data Group, J. Phys. G {\bf 37}, 070521 (2010). http://pdg.lbl.gov/2011/reviews/rpp2011-rev-dark-matter.pdf.

\bibitem{ellisintro}J. Ellis, K.A. Olive, C. Savage, and V.C. Spanos, Phys. Rev. D {\bf 81}, 085004 (2010). 

\bibitem{antares} M. Ageron et al., ANTARES Collaboration,
  Nucl. Inst. and Meth. in Phys. Res. A {\bf 656} (2011) 11-38 [{\tt astro-ph/1104.1607}].

%
%
%
%




%

\bibitem{diffuse} J.A. Aguilar, ANTARES Collaboration, 
 Phys. Lett, {\bf B696} (2011) 16.

\bibitem{ps}  S. Adri\'an-Mart\'{\i}nez et al., ANTARES
Collaboration, Ap. J. Letter {\bf 743} (2011) L14.

\bibitem{monopole}  S. Adri\'an-Mart\'{\i}nez et al., ANTARES
Collaboration, Astropart. Phys. {\bf 35} (2012) 634.

\bibitem{bbfit} J.A. Aguilar, ANTARES Collaboration,
 Astropart. Phys. {\bf 34} (2011) 652.

\bibitem{wimpsim}J. Edsj\"{o}, http://www.physto.se/~edsjo/wimpsim/.

\bibitem{desai}S. Desai et al., Phys. Rev. D {\bf 70}, 083523 (2004).



\bibitem{corsika}D. Heck et al., Report FZKA 6019 (1998), Forschungszentrum Karlsruhe; 
D. Heck and J. Knapp, Report FZKA 6097 (1998), Forschungszentrum Karlsruhe.

\bibitem{bartol}G. Barr et al., Phys. Rev. D {\bf 39} (1989) 3532; V. Agrawal et al., Phys. Rev. D {\bf 53} (1996) 1314.

\bibitem{amram}P. Amram et al., [ANTARES Collaboration], Nucl. Instrum. Meth. A {\bf 484} (2002) 369.


\bibitem{mrf}G.C. Hill, K. Rawlins, Astropart. Phys., 2003, {\bf 19}: pp. 393-402. 
\bibitem{feldmancousins}G.J. Feldman, R.D. Cousins, Phys. Rev., 1998, {\bf D 57}: pp. 3873-3889. 


\bibitem{gondolo}P. Gondolo et al., J. Cosm. and Astropart. Phys., JCAP07, 008 (2004).


\bibitem{superk}T. Tanaka et al., Astrophys. J. {\bf 742}, 78 (2011).

\bibitem{icecube}R. Abbasi et al., Phys. Rev. D {\bf 85}, 042002 (2012).


\bibitem{conrad}F. Tegenfeldt, J. Conrad A NIM A {\bf 539} (2005) 407-413; 
J. Conrad et al., Phys. Rev. D {\bf 67} (2003) 012002; 
J. Conrad (2006) [{\tt astro-ph/0612082v1}].

\bibitem{edsjosdcs}G. Wikstr\"{o}m and J. Edsj\"{o}, J. Cosm. and Astropart. Phys., JCAP04, 009 (2009).

\bibitem{buchmeller}O. Buchmeller et al. [{\tt hep-ph/1207.7315v1}].

\bibitem{baksan}O. Suvorova et al., in proceedings of Dark Side of the Universe, \pos{PoS(DSU2012)042} [{\tt astro-ph/1211.2545}].

\bibitem{picasso}S. Archambault et al., Phys. Lett. B {\bf 711} (2012) 153-161 [{\tt astro-ph/1202.1240}].
\bibitem{kims}H. S. Lee et al., Phys. Rev. Lett. {\bf 108}, 181301 (2012) [{\tt astro-ph/1204.2646}].
\bibitem{simple}M. Felizardo et al., Phys.Rev.Lett. {\bf 108}, 201302 (2012) [{\tt astro-ph/1106.3014}].
\bibitem{coupp}E. Behnke et al., Phys.Rev. D {\bf 86}, 052001 (2012) [{\tt astro-ph/1204.3094}].
\bibitem{xenon}E. Aprile et al. [{\tt astro-ph/1207.5988}].

\end{thebibliography}
\end{document}